\documentclass[prd, twocolumn, superscriptaddress, nofootinbib]{revtex4}
\usepackage{graphicx, epsfig}

\def\cmm2{{\,\rm cm^{-2}}}
\def\cm2{{\,{\rm cm}^2}}
\def\cmm3{{\,{\rm cm}^{-3}}}
\def\gcmm3{{\,{\rm g\,cm^{-3}}}}

\def\fun#1#2{\lower3.6pt\vbox{\baselineskip0pt\lineskip.9pt
  \ialign{$\mathsurround=0pt#1\hfil##\hfil$\crcr#2\crcr\sim\crcr}}}

\newcommand {\ds}{\displaystyle}

\begin{document}

\title{Probing Dark Energy with Supernovae: 
Exploiting Complementarity with the Cosmic Microwave Background}

\author{Joshua A. Frieman\vspace{0.2cm}}
\affiliation{Department of Astronomy \& Astrophysics,
Center for Cosmological Physics,
The University of Chicago, Chicago, IL~~60637-1433\vspace{0.05cm}}
\affiliation{NASA/Fermilab Astrophysics Center
Fermi National Accelerator Laboratory, Batavia, IL~~60510-0500\vspace{0.05cm}}

\author{Dragan Huterer}
\affiliation{Department of Physics
Case Western Reserve University, Cleveland, OH~~44106-7079\vspace{0.05cm}}

\author{Eric V. Linder}
\affiliation{Physics Division, Lawrence Berkeley National Laboratory,
Berkeley, CA~~94720\vspace{0.05cm}}

\author{Michael S. Turner}
\affiliation{Department of Astronomy \& Astrophysics,
Center for Cosmological Physics,
The University of Chicago, Chicago, IL~~60637-1433\vspace{0.05cm}}
\affiliation{NASA/Fermilab Astrophysics Center
Fermi National Accelerator Laboratory, Batavia, IL~~60510-0500\vspace{0.05cm}}
\affiliation{Department of Physics, 
Enrico Fermi Institute,
The University of Chicago, Chicago, IL 60637-1433\vspace{0.05cm}}

\begin{abstract}
\vspace{0.2cm}
A primary goal for cosmology and particle physics over the coming
decade will be to unravel the nature of the dark energy that
drives the accelerated expansion of the Universe. In particular,
determination of the equation-of-state of dark energy, $w\equiv p/\rho$,
and its time variation, $dw/dz$, will
be critical for developing theoretical understanding of the new
physics behind this phenomenon. Type Ia supernovae (SNe) and cosmic microwave
background (CMB) anisotropy are each sensitive to the dark energy
equation-of-state.  SNe alone can determine $w(z)$ with some
precision, while CMB anisotropy alone cannot because of
a strong degeneracy between the matter density $\Omega_M$ and $w$.
However, we show that the Planck CMB mission
can significantly improve the power of a deep SNe survey
to probe $w$ and especially $dw/dz$. Because CMB constraints are
nearly orthogonal to SNe constraints in the $\Omega_M$--$w$ plane,
for constraining $w(z)$ Planck is more useful than precise
determination of $\Omega_M$.
We discuss how the CMB/SNe complementarity impacts strategies
for the redshift distribution of a supernova
survey to determine $w(z)$ and conclude that
a well-designed sample should include a substantial number of
supernovae out to redshifts $z \sim 2$.
\end{abstract}

\maketitle

\section{Introduction}\label{sec:intro}

Recent observations of Type Ia supernovae (SNe) have provided direct
evidence that the Universe is accelerating
\cite{perl99,riess98}, indicating
the existence of a nearly uniform dark-energy component with negative
effective pressure, $w\equiv p/\rho < -1/3$.  Further evidence for
dark energy comes from recent cosmic microwave background (CMB)
anisotropy measurements pointing to a spatially flat, critical density
Universe, with $\Omega_{0} = 1$ \cite{CMB}, combined with a number of 
indications that the matter density $\Omega_M \simeq 0.3$
\cite{omega_m}; the `missing energy' must also have sufficiently
negative pressure in order to allow time for large-scale structure to form
\cite{turner_white}.
Together, these two lines of evidence indicate that dark energy
composes 70\% of the energy density of the Universe and has equation-of-state 
parameter $w < - (0.5 - 0.6)$ \cite{w_constraints}. Determining the nature of
dark energy, in particular its equation-of-state, is a critical
challenge for physics and cosmology.

At present, particle physics theory provides little to no guidance
about the nature of dark energy. A cosmological constant---the
energy associated with the vacuum---is the simplest but not the only
possibility; in this case, $w=-1$ and is time independent, and the
dark energy density is spatially constant. Unfortunately, theory has
yet to provide a consistent description of the vacuum: the energy
density of the vacuum, at most $10^{-10}\,{\rm eV}^4$, is at least 57
orders of magnitude smaller than what one expects from particle
physics---the cosmological constant problem \cite{weinberg}. In recent
years, a number of other dark energy models have been explored, from
slowly rolling, ultra-light scalar fields
to frustrated topological
defects \cite{demodels}. These models predict that $w\not= -1$,
that $w$ may evolve in time, and that there may be small spatial
variations in the dark energy density (of less than a part in $10^5$ on
scales $\sim H_0^{-1}$~\cite{perturbations}). In all models
proposed thus far
dark energy can be characterized by its equation-of-state $w$.
Measuring the present value of $w$ and its time variation
will provide crucial clues to the underlying physics of dark energy.

As far as we know, dark energy can only be probed directly by
cosmological measurements, although it is possible that laboratory
experiments could detect other physical effects associated with
dark energy, e.g., a new long-range force arising from an ultra-light
scalar field \cite{longrange}.
Dark energy affects the expansion rate of the Universe and thereby
influences cosmological observables such as the distance vs.\ redshift,
the linear growth of density perturbations, and the cosmological
volume element (see, e.g.,
\cite{HT}). Standard candles such as Type Ia supernovae offer a direct
means of mapping out distance vs.\ redshift (see, e.g., \cite{WA2002}), while
the CMB anisotropy can be used to accurately determine the distance to
one redshift, the last scattering epoch ($z_{LS}\simeq 1100$).
Because they measure distances at such different redshifts, the SNe
and CMB measurements have complementary degeneracies in the
$\Omega_M$--$\Omega_{\Lambda}$ and $\Omega_M$--$w$
planes \cite{HT,WA2002,CMB+SN}. More recently, Spergel \& Starkman
\cite{SS} have suggested that this complementarity argues for using
supernovae at relatively low redshift, $z\sim 0.4$, to most
efficiently probe dark energy.  In so doing, they used a highly
simplified model which did not consider a spread of SNe in redshift,
systematic error, possible evolution of $w$, or the finite precision
with which planned CMB missions can actually constrain $\Omega_M$ and
$w$.

By including these ``real-world'' effects, this paper clarifies
the complementarity of the CMB and SNe and
explores strategies for best utilizing it in SNe surveys to probe the
properties of dark energy.  We show that dark energy-motivated
supernova surveys should target SNe over a broad range of redshifts
out to $z \sim 2$, and that CMB/SNe complementarity in fact strengthens
the case for deep SNe surveys.

\section{How Supernovae and the CMB Probe Dark Energy}\label{sec:2}

Supernovae and the CMB anisotropy probe dark energy in different ways
and at different epochs.  However, both do so through the effect of
dark energy on the comoving distance vs.\ redshift relation, $r(z)$.
For a spatially flat Universe and constant $w$:
\begin{eqnarray}
H_0r(z) & = & \int_0^z {dz\over H(z)/H_0} \nonumber\\[-0.4cm]
&&\\[0.1cm]
(H/H_0)^2 & = & \Omega_M(1+z)^3 + (1-\Omega_M)(1+z)^{3(1+w)}\nonumber
\end{eqnarray}
where $\Omega_M$ is the present fraction of the energy density
contributed by non-relativistic matter.  This relation is easily
generalized to non-constant $w$ and a curved Universe \cite{HT}; for
notational simplicity we write this and succeeding formulae in terms of
constant $w$, though we generalize them to the evolving case in our analysis.
It is because $H_0r(z)$ depends upon only two
quantities, $\Omega_M$ and $w$, that prior information about $\Omega_M$
(or two independent combinations of $\Omega_M$ and $w$) has such potential
to improve the efficacy of a cosmological probe of dark energy based
upon $H_0 r(z)$.

\begin{figure}[!ht]
\centerline{\psfig{figure=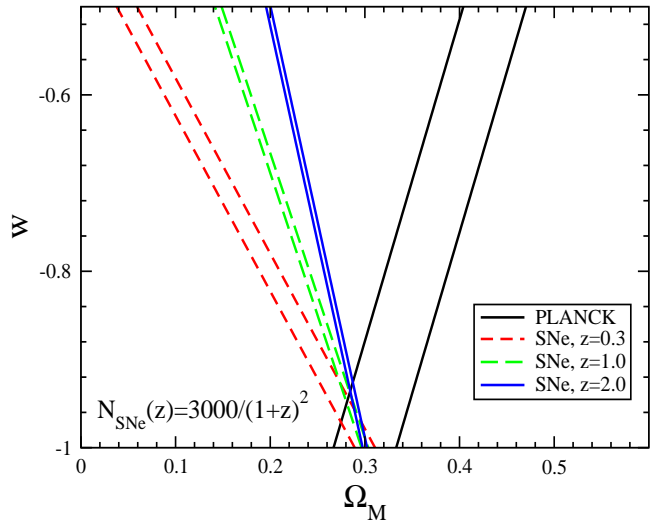, height=3.7in, width=3.0in, angle=-90}}
\caption{68\% CL ``error ellipses'' in the $\Omega_M$--$w$ plane for 3000
SNe all at a single redshift $z=0.3, 1.0$, or $2.0$, and for the
Planck CMB anisotropy measurement (without polarization), assuming a
fiducial model with $\Omega_{M0} = 0.3$ and $w_0=-1$. Because
observations at a single redshift cannot break the parameter
degeneracies, the ellipses do not close.  As expected, the CMB
constraint lies along $\Omega_M \simeq 0.3 + 0.28 (1+w)$.  At higher $z$,
the SNe ellipses become narrower but less orthogonal (complementary)
to the CMB ellipse.  Note, a matter-density prior corresponds
to a vertical stripe, which is less orthogonal to the SNe ellipses
than the CMB ellipse.  This is the basic reason why a CMB prior
is more effective than a matter-density prior.}
\label{fig:Fig1}
\end{figure}

CMB experiments can determine the positions and heights of the
acoustic peaks in the temperature anisotropy angular power spectrum to
high accuracy.  The positions of the acoustic peaks in angular
multipole space depend upon the physical baryon and matter densities
$\Omega_Bh^2$ and $\Omega_Mh^2$, on $\Omega_M$, $w$, and to a lesser
extent other cosmological parameters (e.g., \cite{HT,husug}).
Anisotropy measurements from the Planck \cite{Planck} mission,
planned for launch later in the decade, should
determine the positions of the peaks to better than 0.1\%;
the heights of the peaks will determine $\Omega_Mh^2$ and $\Omega_Bh^2$
(and other cosmological parameters) to roughly percent precision \cite{EHT}.
Together, these measurements should constrain
a combination of $\Omega_M$ and $w$ alone (e.g.,
\cite{HT,SS}) to about 10\% precision. In particular,
in the vicinity of the fiducial values $w_0 = -1$ and
$\Omega_{M0}=0.3$, the combination
\begin{eqnarray}
{\cal D} &\equiv& \Omega_M -0.94\,\Omega_{M0}(w-w_0)\nonumber\\[-0.25cm]
&&\\[-0.15cm]
&\approx& \Omega_M - 0.28(1+w) = 0.3\nonumber
\end{eqnarray}
will be determined to about $\sigma_{\cal D} \simeq \pm
0.03 (\Omega_{M0}/0.3)$ (this result follows directly from
Eq. 18 of Ref.~\cite{HT} by setting $\Delta l/l=
\Delta \Omega_0/\Omega_0 =0$).
The resulting 68\% CL error ellipse in the $\Omega_M$--$w$
plane predicted for Planck is shown in
Fig.~\ref{fig:Fig1}.
Polarization information could in principle improve the precision with which
${\cal D}$ is determined by about 50\% \cite{HETW},
absent problems with foregrounds or the polarization measurements themselves.

The MAP CMB mission \cite{MAP}
currently underway should determine $\cal D$ to a precision
that is about 10 times worse than Planck, assuming temperature
anisotropy information alone. This constraint is too weak to
usefully complement the SNe measurements. However,
if MAP polarization measurements
are successful, this constraint could be improved by about a factor of
two \cite{HETW}; we discuss the potential impact of MAP further
in Sec.~\ref{sec:4-1-3}.

As an aside, we note that the physical baryon and matter densities do
not directly impact the determination of the properties of dark
energy.  Rather, together with other cosmological measurements, they
can be used to determine $\Omega_M$.  In the following Sections we illustrate 
how independent knowledge of $\Omega_M$ can improve the determination of $w$.

Measurements of the energy fluxes and redshifts of Type Ia supernovae
provide an estimate of the luminosity distance as a function of
redshift, $d_L \equiv (1+z)r(z)$. As an example of a supernova survey,
the Supernova/Acceleration Probe (SNAP) \cite{SNAP} is a proposed
space-based telescope to observe $\sim 3000$ SNe Ia out to redshift $z
\sim 1.7$, specifically designed to probe dark energy. To
illustrate the essential principles for such a survey,
though not all the details, we make the simplifying assumption that SNe Ia are
nearly standard candles (after correction for the observed correlation
between light-curve decline rate and peak luminosity \cite{phillips}).
With this assumption, the mean peak energy flux from a
supernova at redshift $z$ is:

\begin{eqnarray}
F(z) & = & {{\cal C} 10^{-0.4M} \over 4\pi d_L^2} = {(10^{10}{\cal
C}/4\pi)10^{-0.4{\cal M}} \over H_0^2d_L^2} = \nonumber\\[-0.2cm]
&&\\
&&
\hspace{-1.2cm}
{(10^{10}{\cal C}/4\pi)10^{-0.4{\cal M}} \over (1+z)^2 \left[ 
\ds {\int_0^z {dz\over 
\sqrt{\Omega_M(1+z)^3 +(1-\Omega_M)(1+z)^{3(1+w)}}}}\right]^2}\nonumber
\end{eqnarray}

\noindent where ${\cal C} = 3.02\times 10^{35}\,{\rm erg\,sec^{-1}}$ is an
unimportant constant, $M$ is the mean absolute peak magnitude of a Type
Ia supernova, and ${\cal M} = M - 5 \log (H_0) + 25$, with distances
measured in Mpc.

It is important to note several things from Eq.\ (3).  First the
energy flux at fixed $H_0 d_L$ depends only upon the combination $\cal
M$ and not upon $M$ and $H_0$ separately.  Thus, the cosmological
parameters $\Omega_M$ and $w$ can be determined by measuring ratios of
fluxes at different redshifts, which are independent of $\cal M$, and
so $\cal M$ is sometimes referred to as a nuisance parameter and can
be easily marginalized over.  Second, since $H_0d_L \rightarrow z$ for
$z\rightarrow 0$, low-redshift supernovae can be used to determine
$\cal M$,
\begin{equation}
z^2F(z) \rightarrow {(10^{10}{\cal C}/4\pi)}10^{-0.4{\cal M}}\ \ {\rm as\ }z\rightarrow 0 .
\end{equation}
For example, a sample of 300 low-redshift supernovae (e.g., as will be
targeted by the Nearby SN Factory \cite{SN_Factory}) could be used to
pin down $\cal M$ to a precision of $\pm (0.01-0.02)$.  Finally, an
absolute calibration of nearby SNe Ia luminosities by another reliable
distance indicator (e.g., using Cepheid variables to determine
distances to galaxies that host SNe Ia
\cite{freedman}) can determine $M$; together, $M$ and $\cal M$ then
fix the Hubble constant, but we emphasize that this is not needed to
probe dark energy.

For a survey of SNe Ia, the likelihood function for the three
parameters the supernova energy flux depends upon is given by
\begin{equation}
{\cal L}_{\rm SNe} (\Omega_M ,w,{\cal M}) \propto \Pi_i \exp
	\left( -\ds{[F_i-F(z_i)]^2\over 2\sigma_i^2} \right)
\end{equation}
where $z_i$ are the redshifts of the supernovae, $F_i$ are their
measured fluxes, and $\sigma_i$ are their measurement
uncertainties (which also includes any random intrinsic spread in peak SNe Ia
luminosities). 

Unlike the CMB, which probes the angular diameter distance at a
single, fixed redshift $z_{LS}$, the efficacy of SNe for determining
$w$ depends upon the redshift distribution of the supernovae. As a
first example, Fig.~\ref{fig:Fig1} shows how well 3000 supernovae at a single
redshift could constrain $\Omega_M$ and $w$, assuming a random flux
error of 0.15 mag per supernova.
Because the sensitivity of the comoving distance $r(z)$ to the dark energy
equation-of-state (e.g., as measured by $dr/dw$) increases with redshift,
the ellipse shrinks for SNe at higher redshift \cite{HT}.

While Fig.~\ref{fig:Fig1} displays important trends, we note that a
realistic survey would not target SNe all at one redshift. Such a
delta-function redshift distribution is very much less than optimal for
constraining $w$ (as we show in Sec.~\ref{sec:4-1}) and would be very
inefficient, since large numbers of discovered SNe would have to be
discarded. More importantly, a broad distribution of SNe redshifts is
crucial for addressing systematic/evolutionary trends in the SNe
population, which must be under control if SNe (or anything else) 
are to be valid probes of dark energy.

In addition, there is much more to studying dark energy than
determining the average value of $w$ in the most efficient manner.
Constraining the time variation of the equation-of-state is critical
for understanding the nature of dark energy. The CMB has no
sensitivity to evolution of $w$; SNe can probe time variation of $w$, and a
broad distribution of SNe redshifts (out to $z \sim 2$) is required to
achieve it, as we show below. In Sec.~\ref{sec:4} we discuss strategies for the
distribution of SNe redshifts and results for some plausible examples.
Finally, determining cosmological parameters (here $\Omega_M$ and
$w$) by two very different techniques has the virtue of providing
consistency checks on the framework of dark energy as well as the
Friedmann-Robertson-Walker cosmology \cite{Li02,Max}.

\section{CMB/SNe Complementarity}\label{sec:3}


Some trends in the CMB/SNe complementarity are illustrated in
Fig.~\ref{fig:Fig1}.  For the fiducial model
($w_0 = -1$, $\Omega_{M0} = 0.3$),
the Planck error ellipse in the $\Omega_M$--$w$
plane is approximately oriented along the line $\Omega_M \simeq 0.3 +
 0.28(1+w)$, as indicated by Eq.\ (2).
By contrast, the error ellipse for 3000 SNe at fixed redshift
has negative slope in this plane;
with increasing redshift it rotates toward $\Omega_M =
{\rm const}$, and its width narrows.  The reason for the rotation is simple:
at high redshift, matter becomes more dynamically important than dark
energy, and the SNe are therefore probing the matter density.  While
the width of the SNe ellipse shrinks with increasing redshift, it becomes less
complementary with the CMB ellipse.  Fig.~\ref{fig:Fig1} also makes it clear why CMB
anisotropy is more complementary than the matter density information:
the matter density prior, which corresponds to a vertical stripe, is
less orthogonal to the SNe ellipse.

To be quantitative, it is useful to write down the joint likelihood function:
\begin{equation}
{\cal L}_{\rm joint} = {\cal L}_{\rm SNe} \times
{\cal L}_{\rm CMB} \times {\cal L}_{\rm other}.
\end{equation}
The CMB likelihood function can be approximated as
\begin{eqnarray}
{\cal L}_{\rm CMB} &=& {\cal L}_{\rm CMB,0} (\Omega_M,w) \times \exp \left[
-\ds{(\rho_B - \rho_{B0})^2\over 2\sigma_{\rho_B}^2} \right]\nonumber\\[-0.3cm]
&&\\[-0.2cm]
&\times& \exp \left[
-\ds{(\rho_M - \rho_{M0})^2\over 2\sigma_{\rho_M}^2} \right]\nonumber
\end{eqnarray}
where
\begin{equation}
{\cal L}_{\rm CMB,0} \ \propto\
\exp \left [\ds{ ({\cal D}-{\cal D}_0)^2\over 2\sigma^2_{\cal D} }\right ] ,
\end{equation}
${\cal D} = \Omega_M - 0.28(1+w)$, ${\cal D}_0 \simeq 0.3$ is the
fiducial value of $\cal D$, $\sigma_{\cal D} \simeq 0.1 {\cal D}_0$ is
the projected accuracy for Planck\footnote{Note that this is merely
illustrative.  In fact we treat ${\cal D}$ by the exact expression for
the distance to the last scattering surface, i.e., Eq.\ (1)
generalized to evolving $w(z)$.},  $\rho_{B} = 1.88\,\Omega_Bh^2 \times
10^{-29}\,{\rm g\,cm^{-3}}$, $\rho_M =1.88 \,\Omega_Mh^2 \times
10^{-29}\,{\rm g\,cm^{-3}}$, $\rho_{B0}$ is the fiducial value of the
baryon density, and $\rho_{M0}$ is the fiducial value of the matter
density.

The accuracy of the CMB constraint in the $\Omega_M$-$w$ plane depends
on three factors: the experiment (and whether polarization information
is included), the parameter set considered, and the presence and
nature of foregrounds. For comparison with SNe without systematics, we
consider the CMB without foregrounds and assuming a moderate set of
eight cosmological parameters and adopt the CMB constraint from
ref.~\cite{Hu_synergy}. Under these assumptions, $\Omega_M h^2$ and
$\Omega_B h^2$ are determined to 1.6\% and 0.8\% respectively for
Planck with temperature information only.  Comparison with SNe with
systematics requires the inclusion of foregrounds, and is obviously
model-dependent. Tegmark et al.~\cite{CMB_sys} have shown that the
accuracy in all cosmological parameters of our interest degrades only
slightly even in the presence of a fairly generous foreground model
(their MID model). Nevertheless, the presence of foregrounds is
expected to degrade the optimistic uncertainties computed without the
systematics (e.g. \cite{Knox}). To account for that, we follow
ref.~\cite{HETW}, where a more generous set of ten parameters is
considered, but without such "luxury parameters" such as running of
the spectral tilt and neutrino mass which were assumed in
ref.~\cite{CMB_sys}. As a result, our model of the CMB with
systematics constrains the quantity $\cal{D}$ 30-40\% worse than the
CMB without systematics.

As noted in Sec.~\ref{sec:2}, the CMB determination of the baryon and matter
densities is not directly useful for constraining dark energy: when
the joint likelihood function is marginalized over the matter and
baryon densities to obtain the one-dimensional probability
distribution for $w$, the integrations over $\Omega_Bh^2$ and
$\Omega_Mh^2$ are trivial.  On the other hand, if we can obtain
information about $M$ (from non-SNe distance measurements) and $\cal
M$ (from low-redshift SNe) and thereby (or otherwise) constrain $H_0$,
then the CMB determination of $\Omega_Mh^2$ constrains $\Omega_M$ as well,
which would directly impact the joint determination of $w$. Of course,
any other external determination of $\Omega_M$ would have the same
effect; later, we will discuss how various $\Omega_M$ priors affect
the determination of $w$.

\begin{figure}[!ht]
\centerline{\psfig{figure=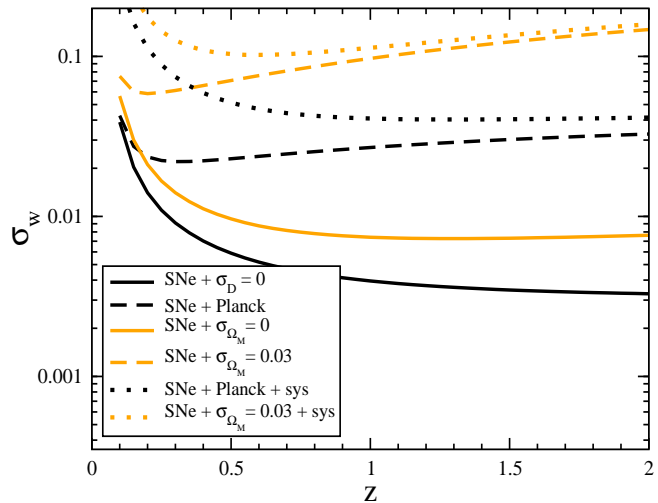, height=3.7in,
	width=3.0in, angle=-90}}
\caption{The predicted 1-$\sigma$ uncertainty in the equation-of-state
parameter $w$ for 3000 SNe {\it at a single redshift}
$z$, with matter density and CMB priors as indicated (and the same
fiducial model as in Fig.~\ref{fig:Fig1}). The dotted curve in each
case includes the effect of a 0.02 mag irreducible systematic error in
measuring the energy flux.  The progression from ``solid to dashed to
dotted'' goes from ``ideal to realistic.'' The legend entries correspond
to order at the very left end, bottom to top.
}
\label{fig:Fig2}
\end{figure}

Assuming no information about $M$ (or equivalently $H_0$), the joint likelihood
function becomes
\begin{equation}
{\cal L}_{\rm joint}(\Omega_M,w) = {\cal L}_{\rm CMB,0}
\times {\cal L}_{\rm SNe}
\end{equation}
{}From this function, we obtain one-dimensional
probability distributions for $w$ by marginalizing over $\Omega_M$.
As a first case, we again assume a baseline sample of 3000 SNe all at
one redshift, with a random flux error of 0.15 mag per supernova.  In
Fig.~\ref{fig:Fig2}, we show the effect of including CMB or $\Omega_M$
information in the determination of the dark energy
equation-of-state, assuming $w = {\rm const}$.  If the CMB measurement
of $\cal D$ is assumed to be ``perfect''
($\sigma_{\cal D} = 0$) as was done in Ref.~\cite{SS},
the predicted $\sigma_w$ drops significantly
with increasing redshift and continues to do so out to $z\approx1.5$.
The effect of a ``perfect'' matter density prior
($\sigma_{\Omega_M}=0$) is similar.  This
qualitative behavior can be understood by referring to
Fig.~\ref{fig:Fig1} and considering the intersection of the CMB line
(now an infinitely thin ellipse) with the SNe ellipses or of a
vertical line (fixed $\Omega_M$) with the SNe ellipses.  The
decreasing width of the SNe ellipses wins out over the decreasing
complementarity at higher redshift.

The qualitative behavior changes, however, when finite precision for
the CMB and matter density measurements is taken into account; as
examples, for
the CMB we use the projected Planck accuracy discussed above, and for the
matter density we assume $\sigma_{\Omega_M} = 0.03$. Not
only is the uncertainty $\sigma_w$ larger in these cases, but it now reaches a
minimum at $z \sim 0.2$ and rises slightly at higher redshift.
For finite widths of the matter density or CMB priors, the
decreasing complementarity now wins out over the decreasing width of
the SNe ellipse with increasing redshift.

Thus far, we have not allowed for systematic error in measuring the
supernova flux at a given redshift.  This means that by measuring a large
number of supernovae at a given redshift, the flux and thereby $r(z)$ can be
determined to arbitrarily high accuracy. In reality, the presence of
residual systematic uncertainty is likely to impose a
floor to improvement.  As a simple model for
irreducible systematic error in the SNe measurements, we assume the
flux error in a specified redshift interval
is given by $\sqrt{ (0.02)^2 + (0.15)^2/N_i}$ mag, where 0.15
mag is the assumed statistical error per SN, 0.02 mag is the
irreducible error,\footnote{In practice, the level of the residual
systematic error depends on survey design, e.g., telescope aperture and
stability, wavelength coverage, observing cadence, point spread function,
seeing, sky background, etc. The systematic error quoted here is based on
the fact that SNAP is specifically designed to achieve 0.02 mag systematic 
error in redshift bins of width $\Delta z = 0.1$.} and $N_i$ is the number of
supernovae observed in that redshift interval.  This model penalizes
observing large numbers of SNe at the same redshift since the
irreducible error adds to the Poisson error: one reaches diminishing
returns for $N_i \sim 100$, at which point the error is only $\sim
20$\% larger than its asymptotic value. While this model is certainly
simplistic, it captures in a straightforward way the essential point:
increasing the number of SNe cannot decrease the measured error in
$H_0r(z_i)$ to arbitrarily small values \cite{basa}.

Figure~\ref{fig:Fig2} illustrates the effect of systematic error.
At redshifts less than about $z\sim 0.5$, systematic
error increases $\sigma_w$ significantly: without the irreducible flux
error, the estimate for $\sigma_w$ was optimistically small because
the flux error was allowed to decrease to a tiny value ($\sim 0.003$
mag).  With systematic flux error included, the predicted error in $w$
from a combined Planck CMB measurement and a hypothetical sample of
3000 SNe (all at redshift $z$) flattens at $z\sim 1$,
with an asymptotic amplitude $\sigma_w \simeq 0.05$.

As noted in Sec.~\ref{sec:2}, realistic survey
would not target supernovae
all at a single redshift, as assumed up to now. Moreover, since the
orientation of the SN error ellipse in the $\Omega_M$--$w$ plane
rotates with $z$ (see Fig.~\ref{fig:Fig1}), a spread of SNe redshifts
helps break the degeneracy between $\Omega_M$ and $w$.  In the next
Section, we consider more realistic strategies for the supernova redshift
distribution to optimally probe dark energy.

\section{Strategies for CMB/SNe Complementarity}\label{sec:4}

\subsection{Optimal}\label{sec:4-1}

The issue of optimal strategies for determining dark energy properties
using SNe in a realistic experiment has been addressed in Refs.~\cite{HT,HL}.
Here, we extend these results to incorporate CMB anisotropy and
other measurements.

\begin{figure}[!ht]
\centerline{\psfig{figure=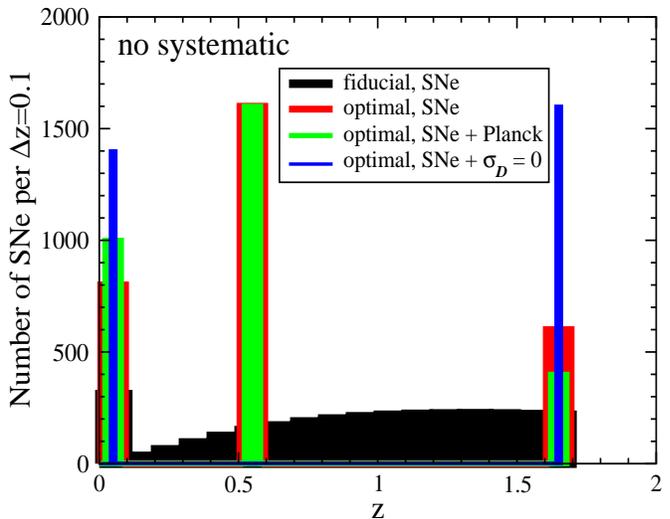, height=3.7in,
	width=3.0in, angle=-90}}
\caption{Optimal redshift distributions in bins of width $\Delta z = 0.1$
for determining $w$ from SNe alone (second widest columns of the
histogram) and with CMB information added (third and fourth
widest). All cases assume $z_{\rm max} = 1.7$ and no systematic error.
The perfect CMB prior (fourth widest, $\sigma_{\cal D} = 0$) is a
``strong'' prior: the optimal distribution comprises two delta
functions; the Planck CMB prior (third widest) is not ``strong'', as
three delta functions remain. For comparison, the black histogram
(widest columns) shows a ``fiducial'' SNAP + SN Factory redshift
distribution with 2812+300 SNe.  }
\label{fig:Fig3}
\end{figure}

\subsubsection{No systematic error}\label{sec:4-1-1}

The optimization problem can be stated as follows: we have three
cosmological parameters ($\cal M$, $\Omega_M$, and $w$; later we will
add a fourth, $dw/dz$); we have ``prior information'' (from the CMB
anisotropy and/or an independent determination of $\Omega_M$); and we
wish to determine the redshift distribution of the SNe which minimizes
the error on $w$, with the constraint that they are confined to the
interval $[0,z_{\rm max}]$. For now, we assume that the total number
of observed SNe is held fixed, and we do not include systematic error
in the SNe measurements. Later we will relax both of these
assumptions.

Huterer \& Turner \cite{HT} showed that for the $N$-parameter problem
with no priors, the optimal redshift distribution comprises $N$ delta
functions, with one at $z=0$, one at $z_{\rm max}$, and the others in
between.  The amplitudes of the delta functions and their positions
relative to $z_{\rm max}$ vary little with the value of $z_{\rm
max}$.\footnote{The optimization can be done with respect to the
errors of the individual parameters or the determinant of the Fisher
matrix (``area of error ellipse'' for the two-parameter problem). The
results in the two cases are similar.  We will minimize
$\sigma_w$ unless otherwise noted.}  Adding a ``strong'' prior on
one, or a combination, of the three parameters reduces the number of
delta functions by one, adding two ``strong'' priors reduces the
number of delta functions by two, and so on.  A ``strong'' prior is
one that constrains one, or a combination, of the three parameters
better than the SNe measurements alone would.  In actuality, this is a
continuous process, with the amplitude of one of the delta functions
going to zero as the quality of the prior improves.  Further, for
smaller $z_{\rm max}$ it is easier to have a ``strong'' prior, since
the SNe constrain the parameters less well.

\begin{figure}[!t]
\centerline{\psfig{figure=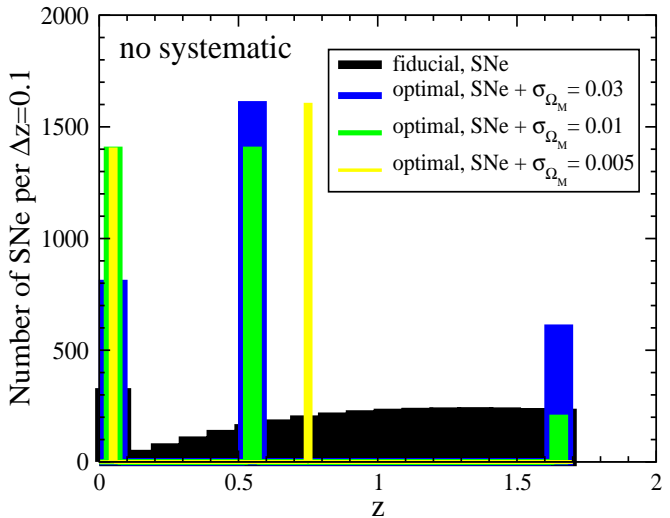, height=3.7in,
	width=3.0in,angle=-90}}
\caption{Same as Fig.~3, except with matter density priors of
$\sigma_{\Omega_M} = 0.03$ (second widest columns), 0.01 (third
widest), and 0.005 (fourth widest). The matter density prior is only
``strong'' for $\sigma_{\Omega_M} \leq 0.005$.  For a strong
matter-density prior, the delta function at $z_{\rm max}$ disappears
because the highest redshift SNe preferentially probe the matter
density.  }
\label{fig:Fig4}
\end{figure}

For illustration we consider a survey of about 3000 SNe with survey depth
$z_{\rm max} = 1.7$. These choices are motivated by the proposed
SNAP survey \cite{SNAP} and thus
provide a useful benchmark (SNAP should obtain 3000 Type Ia
SNe in about two years of observations). Figure~\ref{fig:Fig3} shows
the optimal SNe redshift distribution with no CMB prior, a perfect CMB
prior ($\sigma_{\cal D} = 0$), and the Planck prior (see Sec.~\ref{sec:3}). For
comparison, we also show one of the redshift distributions
currently proposed for SNAP (2812 SNe in the redshift interval
$0.1-1.7$) combined with that for the Nearby SN
Factory (300 SNe at $z< 0.1$).
We see that a ``perfect'' CMB prior is a ``strong'' prior: the optimal SNe
distribution in this case becomes two delta functions, one at $z=0$
and one at $z=z_{\rm max}$.  The Planck prior is not strong: in this
case, three delta functions remain, at $z=0, 0.5$, and 1.7.
Figure~\ref{fig:Fig4} shows the optimal SNe redshift
distribution using $\Omega_M$ instead of CMB priors, with
$\sigma_{\Omega_M} = 0.005, 0.01,$ and 0.03.  The $\Omega_M$ prior
is only ``strong'' for $\sigma_{\Omega_M} \leq 0.005$.

In Figs.~\ref{fig:Fig3} and \ref{fig:Fig4}, the $z \sim 0$ peaks in
the optimal distributions serve mainly to determine $\cal M$. Indeed,
the Nearby SN Factory redshift distribution is strongly peaked at
$z\approx 0.05$, in part for this reason.\footnote{The SN Factory has another
important purpose: the systematic study of Type Ia SNe to better
establish their efficacy as standardizable candles.}  We could have
simply imposed a prior on $\cal M$ instead of including this portion
of the redshift distribution.
 
Finally, it is important to consider how much improvement 
the optimal redshift
distribution actually provides compared to a uniform distribution or the 
SNAP+SN Factory
distribution: for the cases shown in Figs.~\ref{fig:Fig3} and \ref{fig:Fig4}, 
$\sigma_w$ is typically
20\% to 30\% smaller for the optimal distribution.

\subsubsection{Inclusion of systematic error and evolution of $w$}\label{sec:4-1-2}

\begin{figure}[!ht]
\centerline{\psfig{figure=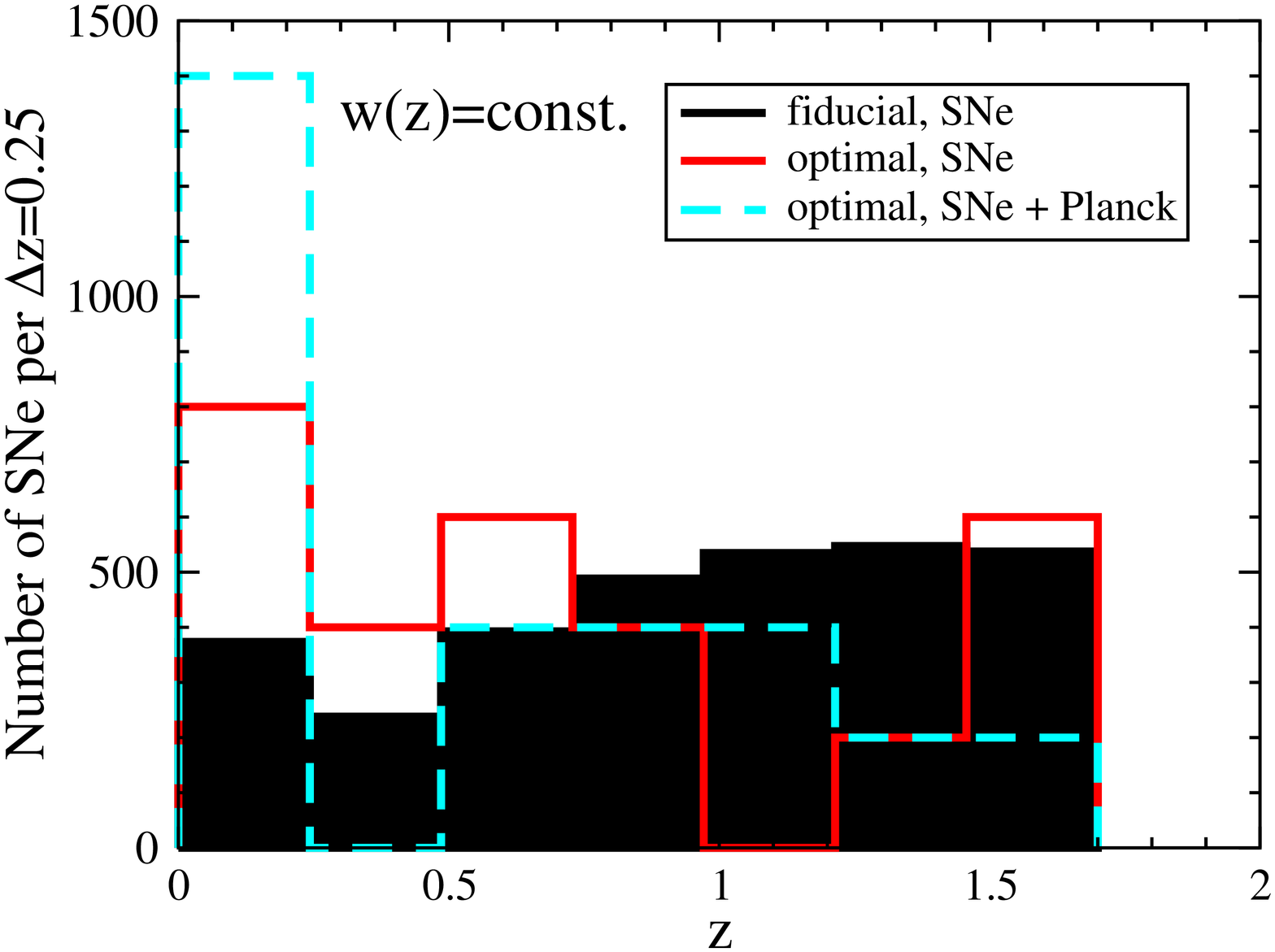, height=3.7in,
 	width=2.8in,angle=-90}}
\vspace{-1cm}
\centerline{\psfig{figure=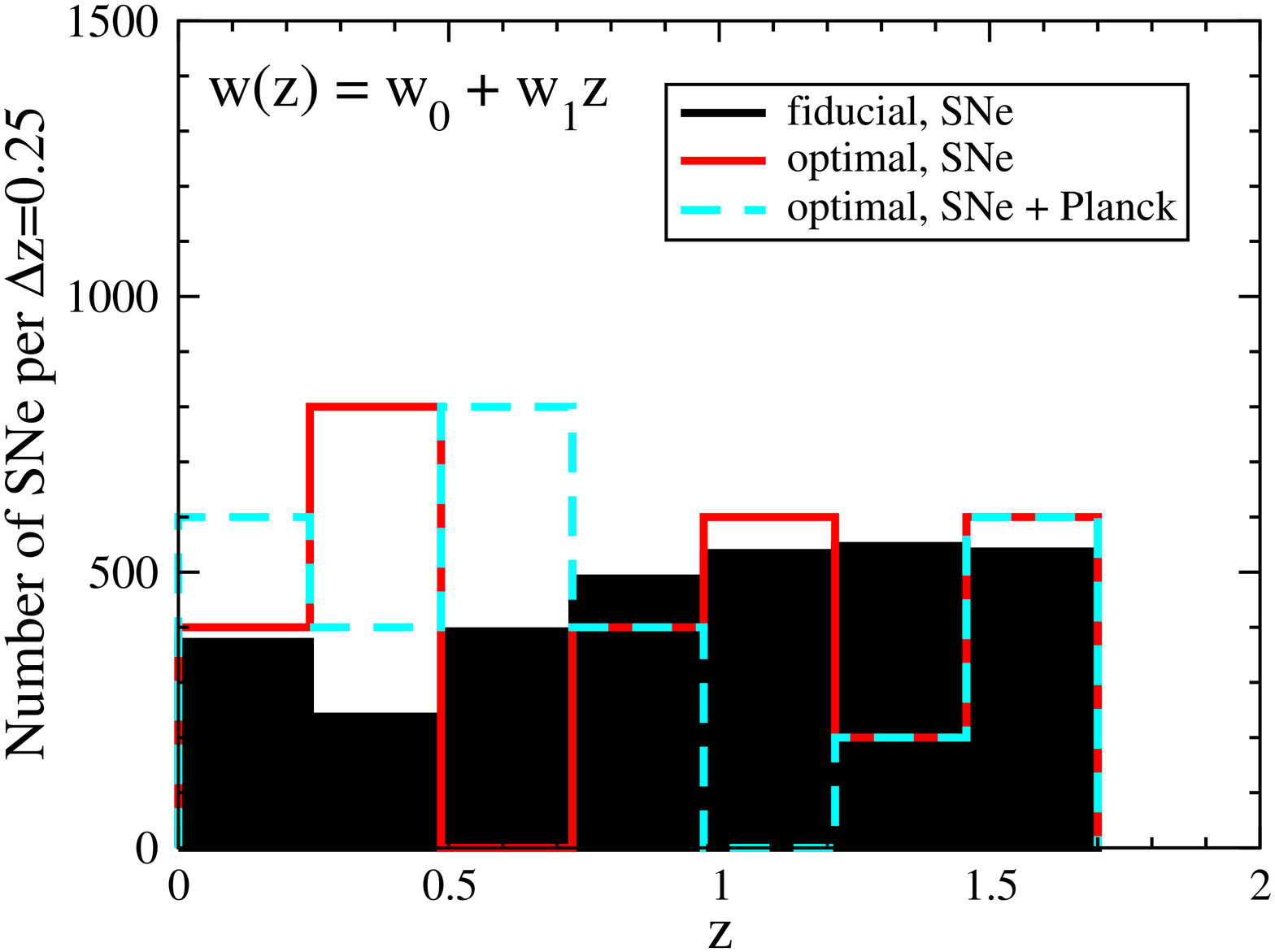, height=3.7in,
	width=2.8in,angle=-90}}
\caption{Optimal redshift distributions for determining $w$ by
3000 SNe measurements alone (solid line) and for 3000 SNe + Planck
CMB measurements (dashed line), with $z_{\rm max} = 1.7$ and
including systematic error. For comparison, the black histogram shows
the fiducial SNAP + SN Factory redshift distribution.  Bins of width
$\Delta z\approx 0.25$ are used solely for numerical convenience.  (a)
Constant $w$; (b) evolving equation-of-state, $w(z)=w_0+w_1z$.  The
optimal distributions are no longer sums of delta-functions when
systematic error is taken into account.}
\label{fig:Fig5}
\end{figure}

\begin{figure}[!ht]
\centerline{\psfig{figure=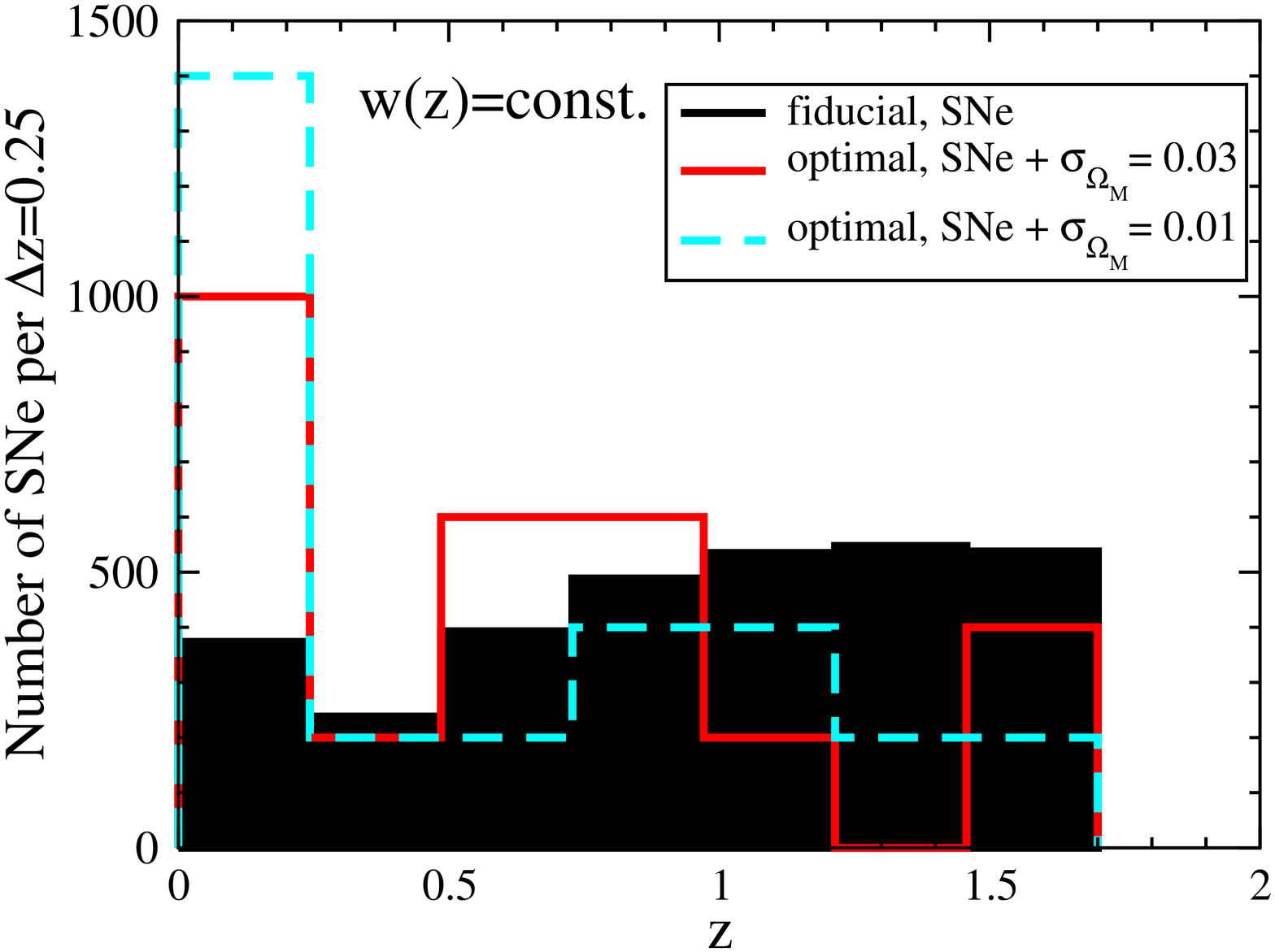, height=3.7in,
	width=2.8in,angle=-90}}
\vspace{-1cm}
\centerline{\psfig{figure=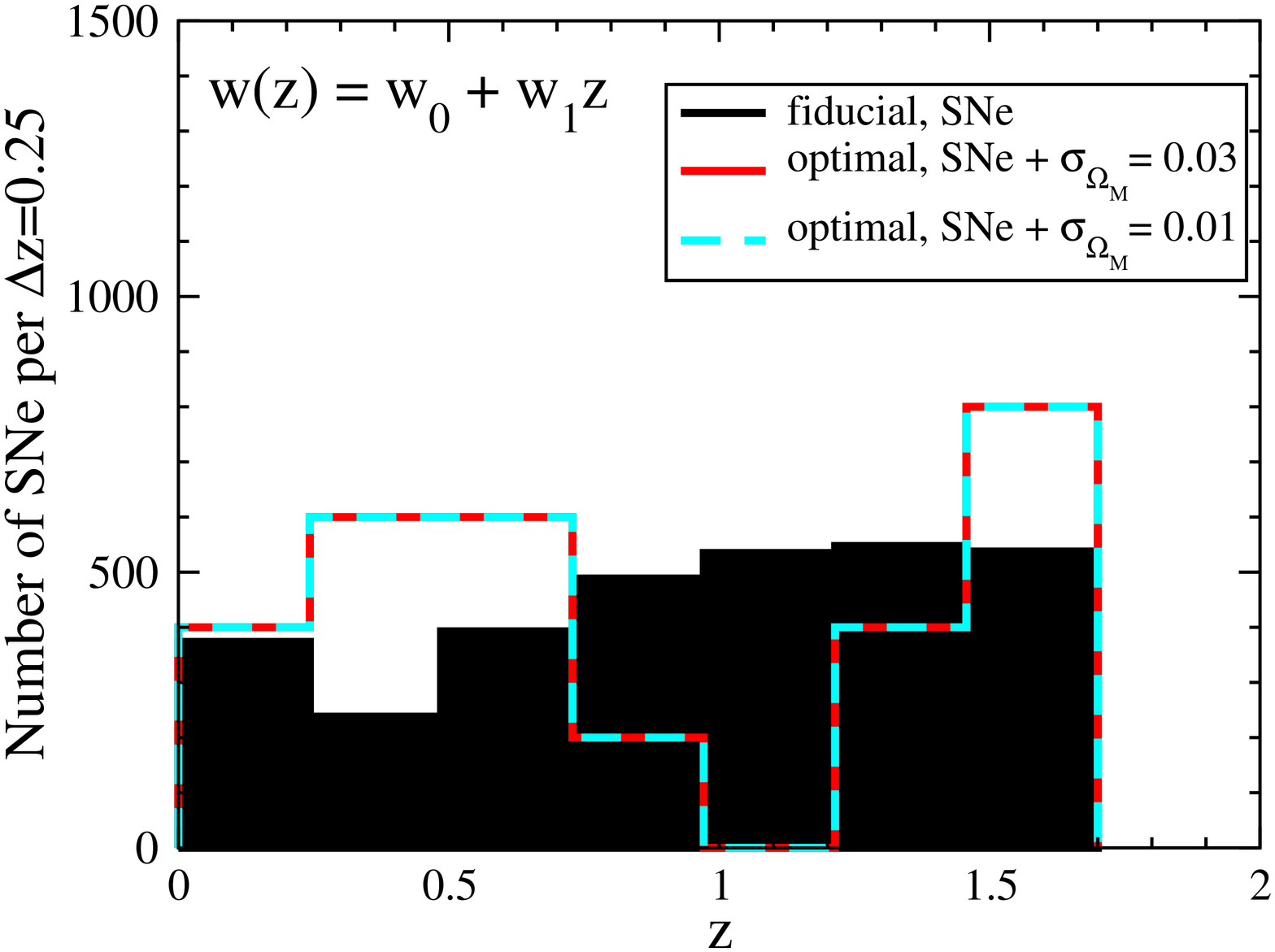, height=3.7in,
	width=2.8in,angle=-90}}
\caption{Same as Fig.~\ref{fig:Fig5}, except for matter-density priors.}
\label{fig:Fig6}
\end{figure}

Now we consider the effect of systematic flux error on the optimal 
SNe redshift distribution. As before, we use the simple model of an
irreducible flux error of 0.02 mag in each redshift interval of width
$\Delta z = 0.1$. We should expect that this will broaden the optimal
distribution, since it is more expedient to spread the remaining SNe
to other redshift bins once the error in a given bin becomes
comparable to the irreducible error. Figs.~\ref{fig:Fig5} and
\ref{fig:Fig6} show the optimal SNe redshift distributions, with and 
without CMB and $\Omega_M$ priors, in the presence of systematic 
errors.  Figs.~5a and 6a show results for the $w={\rm const}$ 
case as before, while
Figs.~\ref{fig:Fig5}b and \ref{fig:Fig6}b 
allow for evolution of the equation-of-state, $w(z) = w_0 +
w_1z$, with $w_1 = dw/dz|_{z=0}$.
Comparison of Figs.~\ref{fig:Fig5}a and \ref{fig:Fig6}a with Figs.~\ref{fig:Fig3} 
and \ref{fig:Fig4} shows that
inclusion of systematic error indeed changes the 
optimal distribution significantly, broadening it to become more uniform.

For the case of constant $w$ (Figs.~\ref{fig:Fig5}a and \ref{fig:Fig6}a),  
the gain in performance for the optimal SNe 
distribution vs.\ a uniform or SNAP+SN Factory distribution is reduced
to only $3-5$\% when systematic errors are included.  
We find that a number of qualitatively different redshift 
distributions yield essentially the same value of $\sigma_w$. 
In particular, in this case $\sigma_w$ is relatively insensitive 
to $z_{\rm max}$: 
there exist distributions with no SNe at $z>1$ which yield $\sigma_w$
only 3\% larger than the optimal value (see also Fig.~\ref{fig:Fig7} below).

The situation is markedly different if we allow for time
variation in the equation-of-state.  In Figs.~\ref{fig:Fig5}b and 
\ref{fig:Fig6}b, we
show the distributions that minimize $\sigma_{w_1}$
(the results are almost identical if $\sigma_{w_0}$ is
minimized instead). In the presence of CMB or matter
density priors, the optimal
distributions now include larger numbers of SNe at high
redshift.   Furthermore, SNe in the high-redshift range $1 < z < 1.7$
are crucial for precision constraints to $w_1$,
even in the presence of a strong prior.  For example,
as $z_{\rm max}$ increases from 1 to 1.7,
$\sigma_{w_1}$ decreases by more than a factor of two,
cf. Fig.~\ref{fig:Fig9}.

\subsubsection{Gains from complementarity}\label{sec:4-1-3}

\begin{figure}[!ht]
\centerline{\psfig{figure=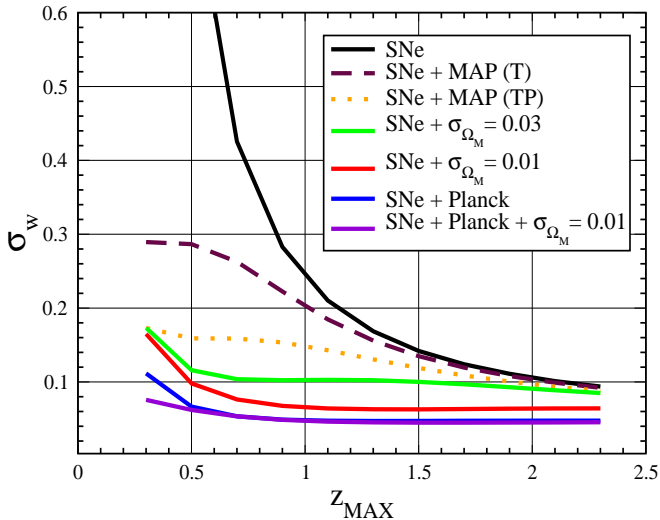, height=3.7in,
	width=3.0in,angle=-90}}
\caption{The predicted $\sigma_w$ vs. SNe survey depth for a combined
set of experiments, ordered from top to bottom: (a) SNe only, (b) SNe
+ MAP (temperature only), (c) SNe + MAP (temperature and
polarization), (d) SNe + ($\sigma_{\Omega_M} = 0.03$), (e) SNe +
($\sigma_{\Omega_M} = 0.01$), (f) SNe + Planck, and (g) SNe + Planck +
($\sigma_{\Omega_M} = 0.01$).  In all cases, we assume the scaled SNAP
+ SN Factory redshift distribution and an irreducible systematic error
in flux measurements of 0.02 mag in redshift bins $\Delta z = 0.1$. }
\label{fig:Fig7}
\end{figure}

The preceding analysis shows that, for fixed $z_{\rm max}$, the error
on $w$ is only weakly dependent on the SNe redshift distribution: in
the presence of systematic error, distributions which are broadly
spread over the range $0<z<z_{\rm max}$ differ only slightly in their
performance. Therefore the chief determinant of the error is $z_{\rm
max}$ itself, and we now address how the efficacy of SNe with
complementary information depends on this maximum redshift. In
Fig.~\ref{fig:Fig7}, we show the effect of various CMB and matter
density priors on the predicted value of $\sigma_w$ vs.\ $z_{\rm max}$,
assuming $w={\rm const}$,
with systematic error modeled as before and assuming a scaled version of
the SNAP + SN Factory
distribution of redshifts.\footnote{When varying $z_{\rm max}$ from
its fiducial value of 1.7, we
truncate the fiducial SNAP distribution at the new $z_{\rm max}$ and
scale it to preserve the total of 2812 SNe. The SN Factory
distribution is then added unchanged -- 300 SNe in the lowest redshift
bin.}  (As noted above, the optimal redshift distribution with
the same value of $z_{\rm max}$ would yield
only slightly smaller $\sigma_w$.)  Figure~\ref{fig:Fig7} also
includes the error on $w$ for the case of no CMB prior or knowledge of
the matter density (black curve).

The primary effect of incorporating additional information, from
either the CMB or the matter density, is to dramatically decrease
$\sigma_w$ at redshifts less than one and thereby lessen the
dependence of $\sigma_w$ on $z_{\rm max}$.  With SNe only, $\sigma_w$
decreases from 0.8 to 0.15 as $z_{\rm max}$ is increased from 0.5 to
1.5.  With the Planck or matter density prior, $\sigma_w$ decreases
less rapidly and levels off at $z\sim 1$.
Note that the Planck prior is more
effective than either matter density prior shown.  Even combining a
$\sigma_{\Omega_M} = 0.01$ prior with Planck provides little
improvement over the Planck prior alone.  Although an independent
determination of $\Omega_M$ to $\pm 0.03$ can substantially improve
the precision with which $w$ can be determined if $z_{\rm max} \leq
1.5$ \cite{HT}, the Planck CMB prior by itself does better by a
factor of two.

\begin{figure}[!ht]
\centerline{\psfig{figure=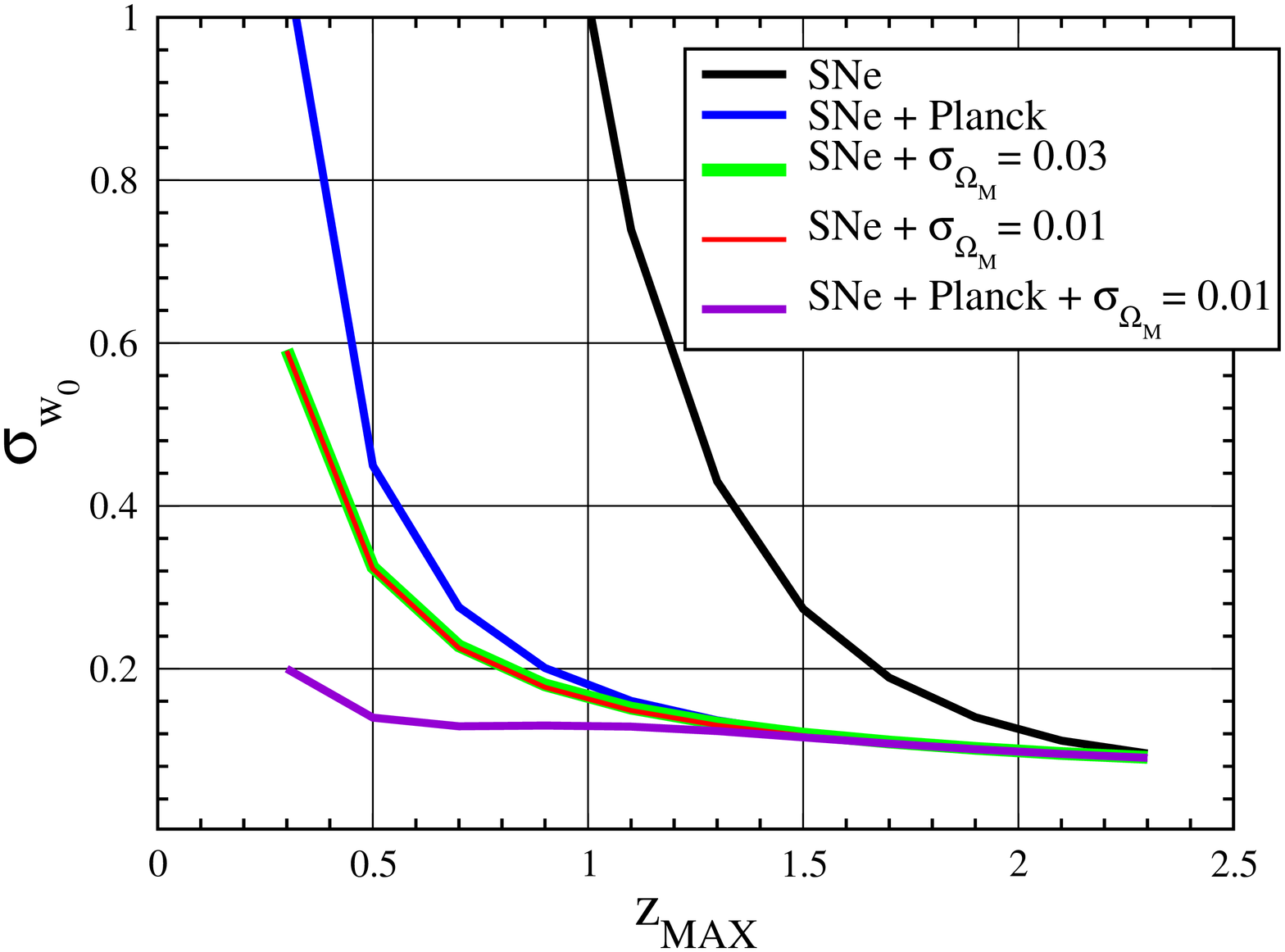, height=3.7in,
	width=3.0in,angle=-90}}
\caption{Same as Fig.~\ref{fig:Fig7}, but for $w_0$, where $w(z)
=w_0 + w_1z$. The curves for SNe + MAP are not shown.
}
\label{fig:Fig8}
\end{figure}

\begin{figure}[!ht]
\centerline{\psfig{figure=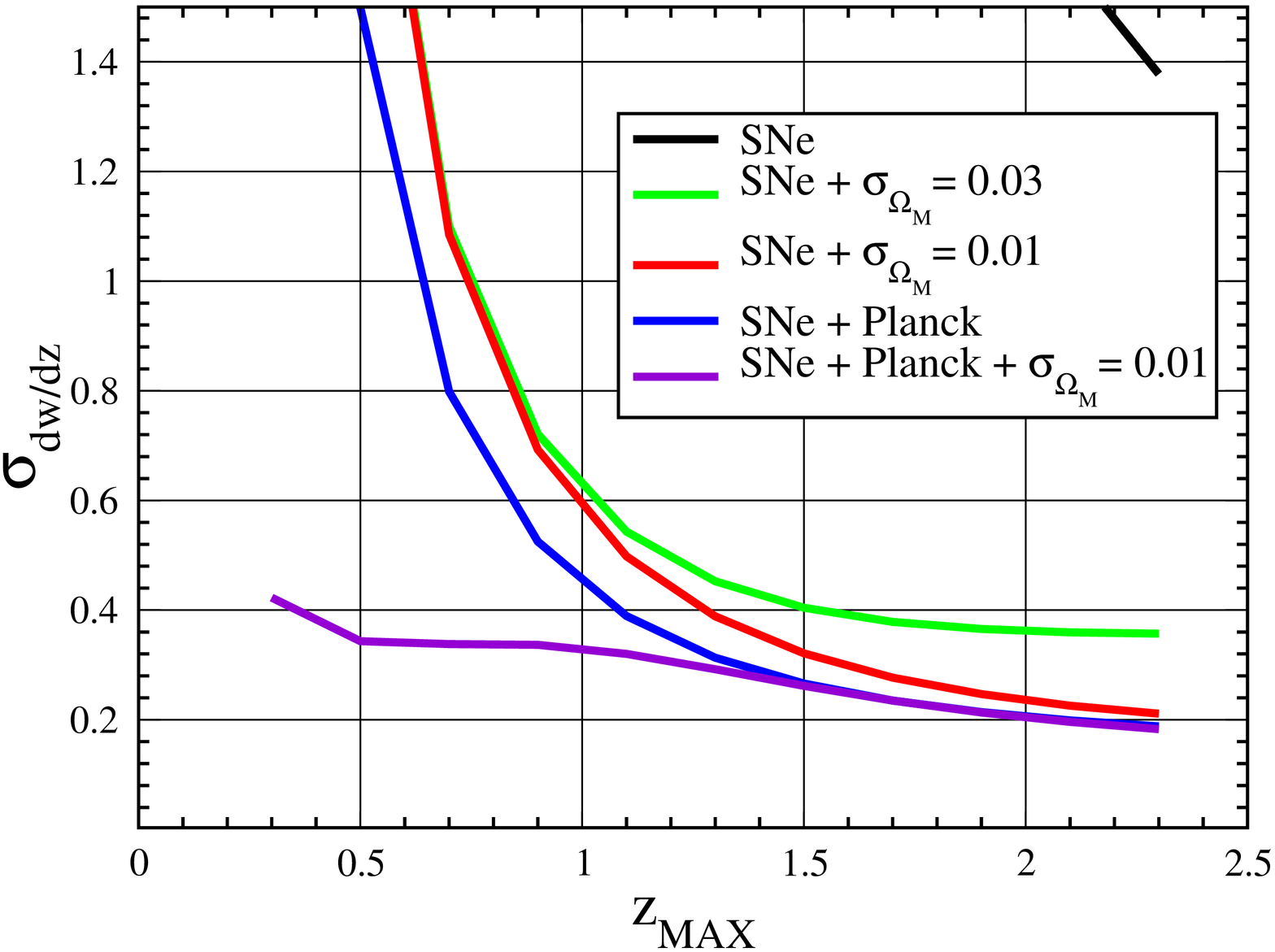, height=3.7in,
	width=3.0in,angle=-90}}
\caption{Same as Fig.~\ref{fig:Fig7}, but for $w_1 = (dw/dz)|_{z=0}$,
where $w(z) = w_0 + w_1z$. The curves for SNe + MAP are not shown.
}
\label{fig:Fig9}
\end{figure}

As mentioned at the end of Sec.~\ref{sec:2}, time
variation in the equation-of-state is generically expected and is a
potentially important discriminator between dark energy models.
Allowing for evolution, with
$w(z) = w_0 + w_1 z$,\footnote{As discussed in Ref.~\cite{HT},
the exact form chosen for the parameterization is not essential.}
there are now four parameters to determine:  ${\cal M}, \Omega_M, w_0,$, and $w_1$.
As Figs.~\ref{fig:Fig8} and \ref{fig:Fig9} illustrate,
without an additional prior, SNe have little leverage on $w_0$ and
$w_1$ \cite{HT,Maoretal}.  An independent determination of the matter
density to $\pm 0.03$ -- not much more stringent than already achieved: 0.04
\cite{omega_m} -- would allow $w_0$ and $w_1$ to be determined to
precision of about $\pm 0.1$ and $\pm 0.35$ for $z_{\rm max} \sim
1.7$ \cite{HT}.  The Planck prior is just as good as a
$\sigma_{\Omega_M} = 0.03$ matter density prior for $w_0$ (if $z_{\rm
max} \geq 1$) and better for $w_1$.  Note that the improvement with
survey depth in $\sigma_{w_1}$ (and to a lesser extent $\sigma_{w_0}$)
continues out to $z_{\rm max}=2$ in all cases. That is, even in the
presence of complementary information from the CMB or the matter
density, a SNe survey aimed at detecting and constraining the
evolution of the dark energy equation-of-state should extend out to high
redshift, $z_{\rm max} \sim 1.5 - 2$.

Thus far, our discussion of CMB anisotropy has been confined
to the Planck mission.   It is also worth
considering what can be learned from the ongoing MAP experiment.  
As noted in Sec.~\ref{sec:2}, with temperature anisotropy measurements alone,
MAP can determine $\cal D$
about 10 times less accurately than Planck, $\sigma_{\cal D} \simeq 0.3$.
In this case, MAP provides a far less useful prior than 
the matter density prior $\sigma_{\Omega_M} =0.03$
(about a factor of two worse for $\sigma_w$), cf. Fig.~\ref{fig:Fig7}.
Even if MAP can achieve its full polarization capability (a factor
of two improvement in $\sigma_{\cal D}$~\cite{HETW}), 
a MAP prior is still not as good as
the matter density prior $\sigma_{\Omega_M} =0.03$.  Moreover,
mapping the polarization anisotropy on large angular scales --- where it helps
determine $w$ indirectly, by imposing an upper limit to the ionization
optical depth $\tau$ --- will be difficult in the presence of polarized
synchrotron radiation from the Galaxy.  
Finally, we mention that while polarization measurements
also have the potential to improve the Planck determination of $\cal D$ (by
about 50\%), this only improves the joint SNe/CMB determination of
$w$ by about 15\%.  The reason is simple:  it is the width of the SNe
error ellipse that controls $\sigma_w$.

\subsection{Resource limited}\label{sec:4-2}

In the analysis so far, we have assumed a fixed total number of
observed supernovae, $N_{SN} = 3112$. However, the resources required
to discover and follow up a supernova depend in general upon its
redshift. Thus, an important but more complicated problem involves the
optimization of the determination of dark energy parameters with
fixed total resources.  Actually determining what these fixed
resources are (e.g., discovery time, follow-up time, spectroscopy
time) and how much each supernova `costs' is beyond the scope of this
paper (relevant ongoing studies can be found at
\cite{SNAP}). We note that these costs will depend in detail upon a
variety of technical factors: telescope aperture, pixel size and
number, CCD quantum efficiency, sky brightness, atmospheric seeing
(for ground-based observations), required signal to noise, etc.

\begin{figure}[!ht]
\centerline{\psfig{figure=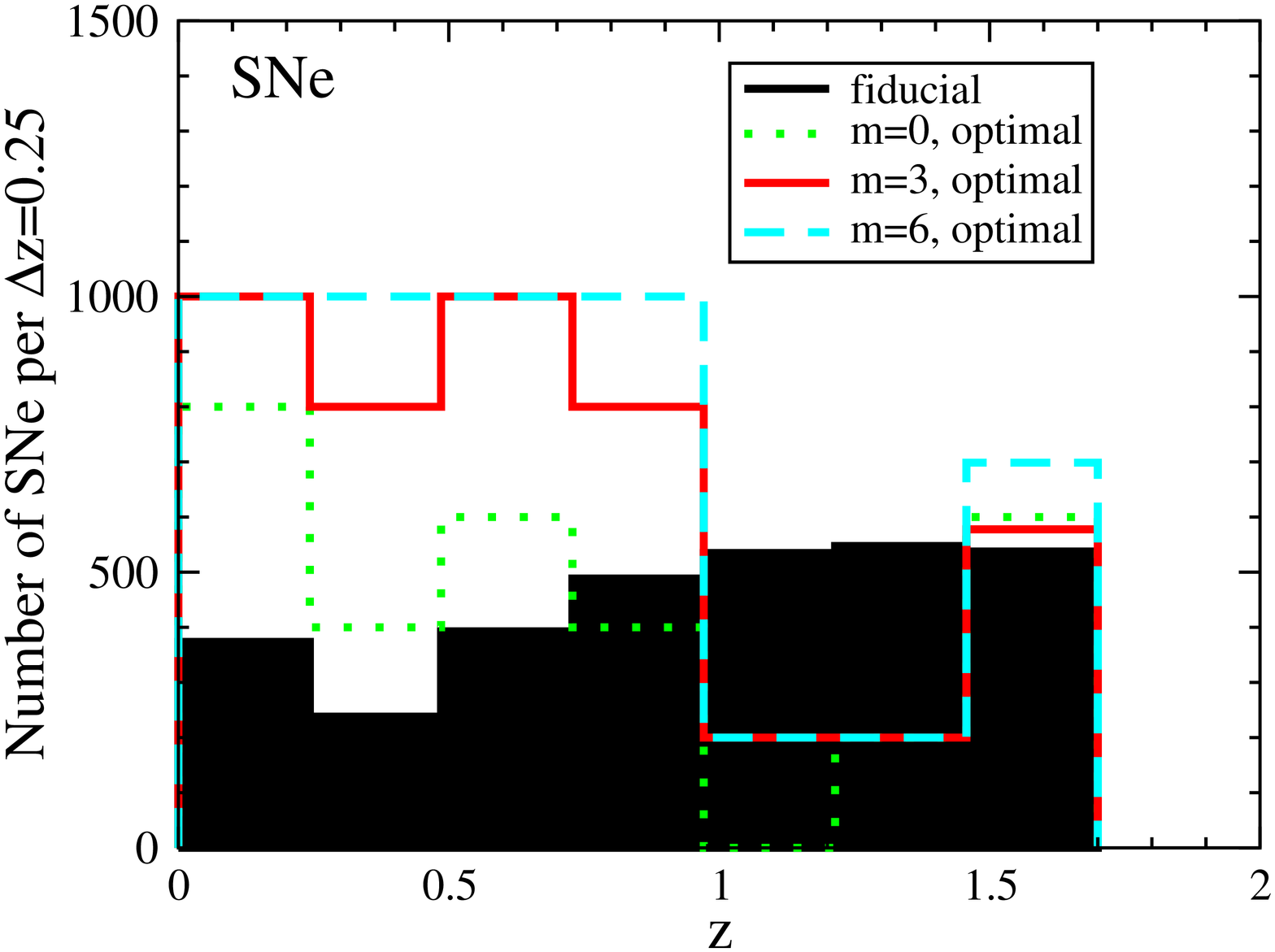, height=3.7in,
	width=2.8in,angle=-90}}
\vspace{-1cm}
\centerline{\psfig{figure=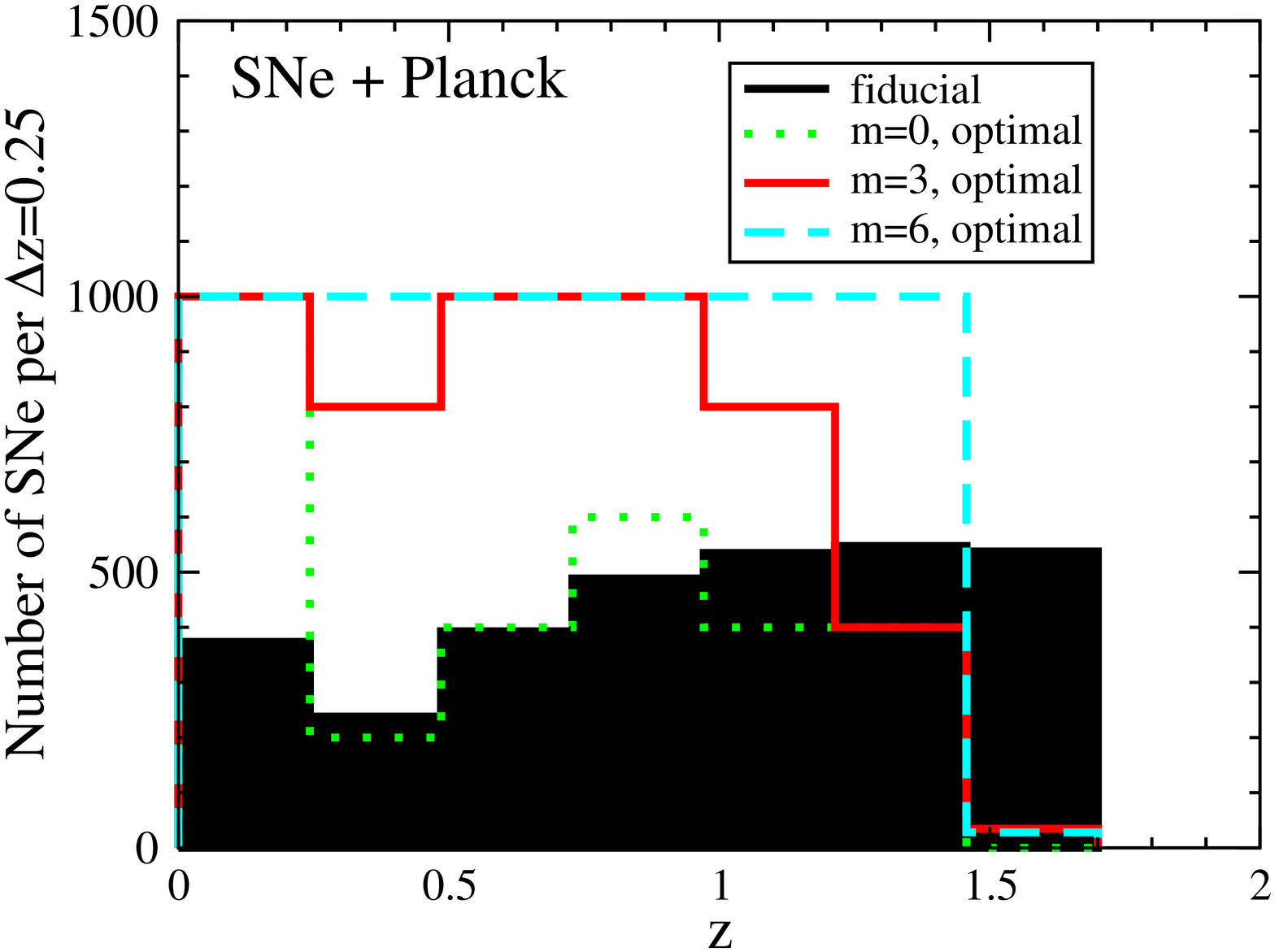, height=3.70in,
	width=2.8in,angle=-90}}
\caption{The resource-optimized redshift distributions for determining
(constant) $w$ by (a) SNe measurements alone and (b) SNe + Planck,
including systematic errors, assuming the cost per
supernova scales as $(1+z)^m$, for $m=0,3,6$. The fiducial SNAP + SN
Factory distribution is shown for comparison.}
\label{fig:Fig10}
\end{figure}

\begin{figure}[!ht]
\centerline{\psfig{figure=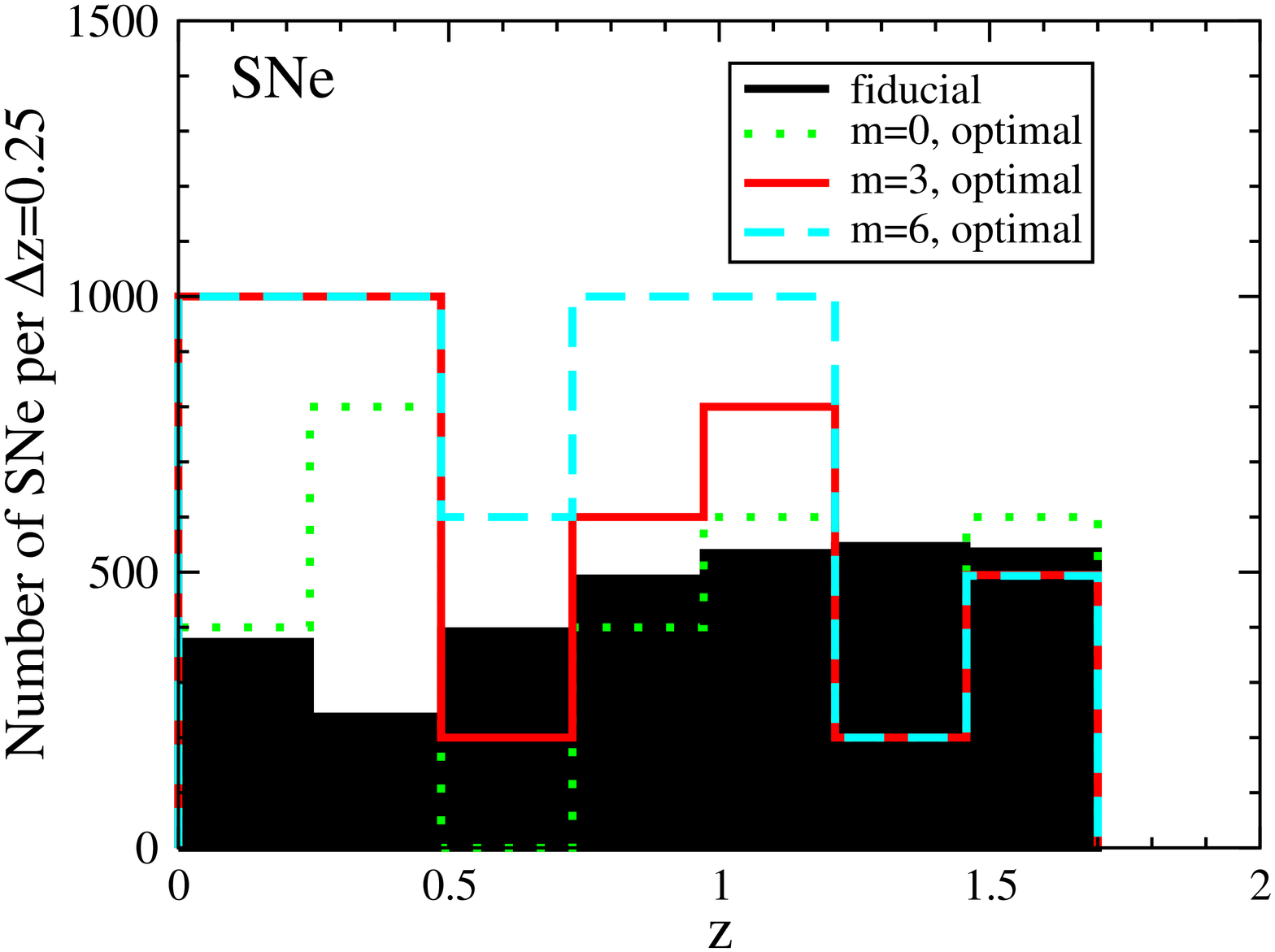, height=3.7in,
	width=2.8in,angle=-90}}
\vspace{-1cm}
\centerline{\psfig{figure=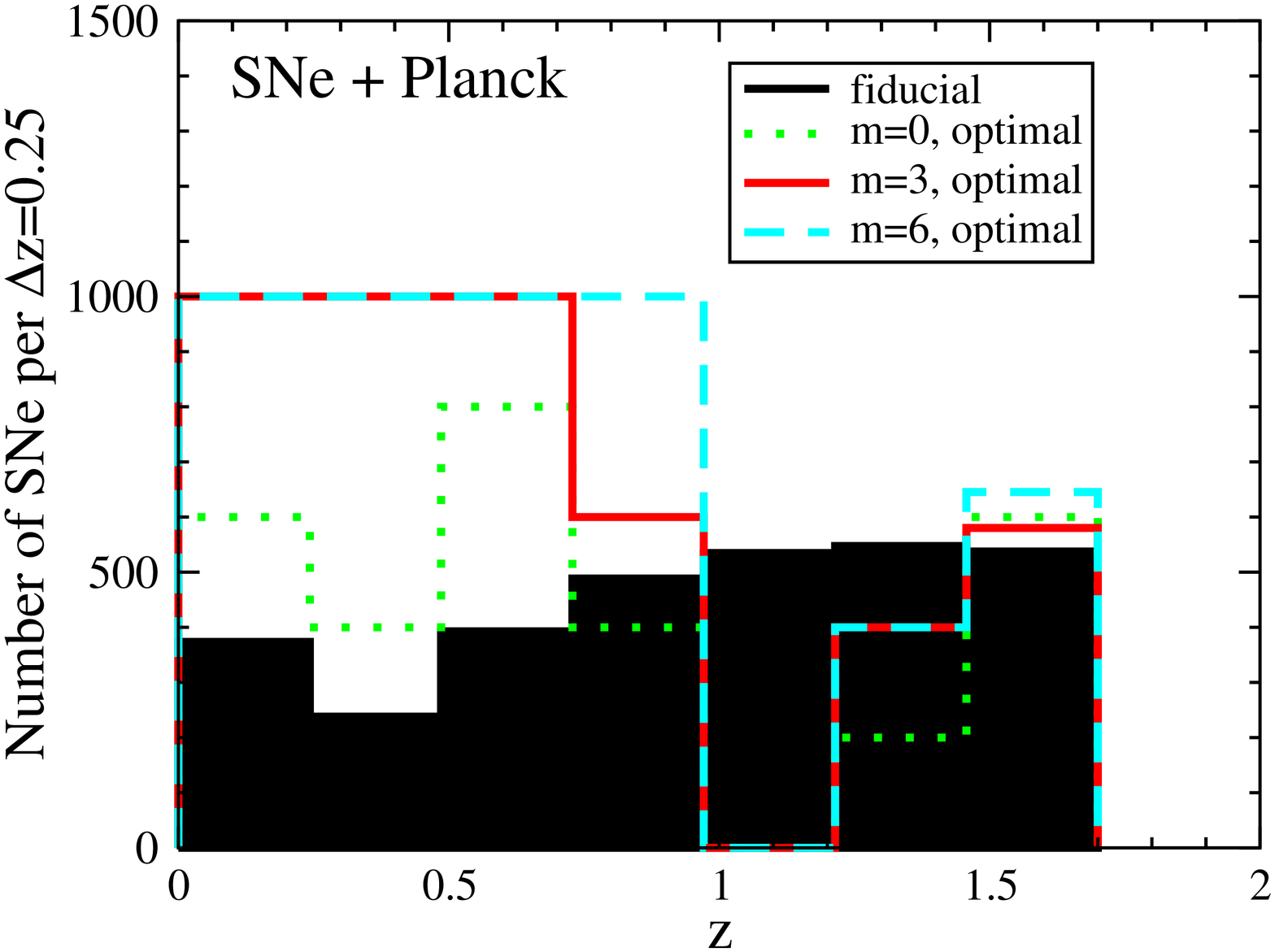,
	width=2.8in,angle=-90}}
\caption{Same as Fig.~\ref{fig:Fig10}, but with $w(z)=w_0+w_1\,z$.
}
\label{fig:Fig11}
\end{figure}

As a highly simplified model, let the normalized cost of each
supernova observed at redshift $z$ be $(1+z)^m$, so that the total
cost of a survey that follows up $N$ supernovae is
 $\sum_{i=1}^N (1+z_i)^m$. The problem is to find the optimal SNe
redshift distribution for fixed total resources (total cost)
$R$. For SNAP, the observing time cost
for spectroscopy or photometry per supernova is estimated to scale as
$(1+z)^6$ for fixed signal to noise \cite{SNAP}.  In the case of wide
field, multiplexing photometry that SNAP is designed
for, simultaneously discovering and following up supernovae by
repeatedly sweeping the same field could reduce this
by a large factor. To span the plausible
range of cost functions, we show results for $m=0, 3$, and 6.

To fix the total resources $R$, we assume that there are
sufficient resources to carry out a survey of 3112 SNe with the
fiducial SNAP + SN Factory redshift distribution shown, e.g., in
Fig.~\ref{fig:Fig3}.  That is, for a
given value of $m$, we fix $R$ by computing the total cost of the
fiducial SNAP + SN Factory redshift distribution.  Then we find the SN
redshift distribution that minimizes $\sigma_w$ within the resource
constraint, i.e., for the same value of $R$. If we place no
upper bound on the number of SNe per redshift bin, the number of SNe
at low redshifts would be driven to huge values as $m$ is increased.
Clearly a distribution with many thousands of SNe in any
redshift bin is not experimentally
realistic, and the systematic error makes this
an unwise choice: the gains in terms of reduced $\sigma_w$ are
negligible once the number of SNe per bin goes much above 100.
We therefore impose the further
constraint that the number of SNe per redshift bin of width
0.25 not exceed (a very generous) 1000.

The results for $m=0, 3$, and 6 are shown in Figs.~\ref{fig:Fig10} and
\ref{fig:Fig11}, again for $z_{\rm max} = 1.7$, the same model for
irreducible systematic error as above, and either no prior from
the CMB (Figs.~\ref{fig:Fig10}a,~\ref{fig:Fig11}a) or the Planck prior 
(Figs.~\ref{fig:Fig10}b,~\ref{fig:Fig11}b).
In Fig.~\ref{fig:Fig10}, we assume constant $w$, while in
Fig.~\ref{fig:Fig11} $w$ can evolve.
We note that the performance of the optimal distribution in minimizing
$\sigma_w$ (or $\sigma_{w_1}$) is only 2 to 10\% better than the SNAP + SN
Factory distribution in all cases.

Consider first the constant $w$ case. Figure~\ref{fig:Fig10}
shows that, as $m$ increases, SNe start filling up the
lower redshift bins to the maximum allowed number; this continues
until the resource limit is reached. While this is strictly true for
the Planck prior, with no prior a
significant fraction of SNe remain in the highest redshift bin. This
behavior can be understood simply: without any priors, the high
redshift SNe are crucial for breaking the degeneracy between
$\Omega_M$ and $w$ (see Fig.~\ref{fig:Fig7});
the addition of the Planck prior partially breaks this
degeneracy, and the number of SNe in the highest redshift bin therefore
decreases.

The case of evolving $w$ is qualitatively similar, with one important
difference: the high-$z$ subsample of SNe is always present in the optimal
distribution, regardless of the prior or the value of $m$. As
Fig.~\ref{fig:Fig11} shows, the highest redshift bin always has a significant
number of SNe ($\simeq 500$), even for $m=6$, when their cost is large.

Although the exact optimal distribution for a given value of $m$, and the
corresponding values of $\sigma_w$ and $\sigma_{w_1}$, will
depend in practice on details of the
optimization --- the number of redshift bins and the maximum number of SNe
allowed per bin --- some clear trends emerge from this analysis.
While the lower redshift
bins become relatively more populated in the optimal distributions
(reflecting the lower cost of low-redshift SNe), the importance of
high redshift supernovae remains: in {\it all} cases, at least 800 SNe are
at redshifts $z > 1$. For the constant $w$ case with no Planck prior,
or for evolving $w$ regardless of prior, these high-redshift SNe
are crucial to making the error on $w$ small enough to be useful.

Clearly we have just scratched the surface with regard to resource-limited
optimization; to proceed further, one would need a much
more quantitative description of the resources available and the systematics.

\section{Summary and Conclusions}\label{sec:concl}

Unraveling the nature of dark energy is one of the outstanding
challenges in physics and astronomy.  Determining its properties is
critical to understanding the Universe and its destiny and may 
shed light on the 
fundamental nature of the quantum vacuum and perhaps even of space-time.
Type Ia supernovae and CMB anisotropy can both 
probe the dark energy equation-of-state $w$, and we have
explored in detail the synergy between the 
two.  With the MAP mission in progress, the Planck mission
slated for launch in 2007, and the design of dedicated
SN surveys now underway, such a study is very timely.

CMB anisotropy alone cannot tightly constrain the properties of dark
energy because of a strong degeneracy between the average
equation-of-state and the matter density.  SNe can probe $w$ with a
precision that improves significantly with knowledge of the matter
density, because $H_0r(z)$ depends only upon $w$ and $\Omega_M$.  A
key result of this paper is that CMB anisotropy measurements by the
upcoming Planck mission have even more potential for improving the
ability of SNe to probe dark energy. The reason is simple: in the
$\Omega_M$--$w$ plane (Fig.~\ref{fig:Fig1}), the CMB constraint is
more complementary to the SNe constraint than is determination of
$\Omega_M$.

Compared to the matter density prior $\sigma_{\Omega_M} = 0.03$,
Planck CMB data reduce the predicted error $\sigma_w$ (under the
assumption of constant $w$) by about a factor of two
(Fig.~\ref{fig:Fig7}). In probing possible variation of $w$ with
redshift, the Planck prior is also significantly better than the same
matter density prior (Fig.~\ref{fig:Fig9}).  Given the concern
expressed by some (e.g., \cite{Maoretal}) that a precise measurement
of the matter density independent of dark energy properties may be
difficult, this is good news.  On the other hand, we find that even if
MAP can successfully measure polarization on large scales, its
potential for complementarity with SNe falls short of that for Planck
and is not as good as the $\sigma_{\Omega_M}=0.03$ matter density
prior.

We have also explored how the SNe determination of the dark energy
equation-of-state, with or without prior information from the CMB or
the matter density, depends upon the redshift distribution of the
survey, including the effects of systematic error and a realistic
spread of SNe redshifts. For either constant or evolving $w$, the
optimal strategy calls for significant numbers of SNe above redshift
$z\sim 1$.  For the constant $w$ case with no Planck prior, or for
evolving $w$ regardless of prior, these high-redshift SNe are
necessary for achieving $\sigma_w < 0.1$. Observing substantial
numbers of SNe at these high redshifts also provides the only hope of
probing time evolution of the equation-of-state with reasonable
precision.  Moreover, the improvement in $\sigma_{dw/dz}$ continues to
high redshift: $\sigma_{dw/dz}$ falls by more than a factor of two
when $z_{\rm max}$ increases from $1$ to $2$ (Fig.~\ref{fig:Fig9}).
Since we currently have no prior information about (or consensus
physical models which significantly constrain) the time variation of
$w$, the design of a SNe survey aimed at probing dark energy should
take into account the possibility that $w$ evolves. These conclusions
about the need for high-redshift supernovae do not change
significantly if we consider a hypothetical survey for which resources
are constrained and a redshift-dependent cost is assigned to each
supernova.

Ref.~\cite{SS} raised the question whether a shallow SNe survey 
is better than a deep one in determining 
the dark energy equation-of-state, given prior
knowledge from the CMB. 
Our results indicate that it is not, once the SNe and 
CMB experiments are realistically modelled. 
On the contrary, CMB/SNe complementarity strengthens the
case for a deep SNe survey that extends to redshift $z \sim 2$.

\begin{acknowledgments}
This work was supported by the DoE (at Chicago, Fermilab, LBL, and
CWRU), NASA (at Fermilab by grant NAG 5-7092), and the NSF Center for
Cosmological Physics at Chicago.  EL would like to thank Ramon Miquel
and Nick Mostek for their help with computations. DH thanks Wayne Hu for
conversations regarding the MAP and Planck CMB missions.
\end{acknowledgments}

\end{document}